\documentstyle[12pt]{article}
\begin{document}
\vspace{1cm}
\centerline{\large{Deltaron Dibaryon Structure in Chiral SU(3) Quark Model}
\footnote{This work was partly supported by the National
Natural Science Foundation of China }}
\vspace{1cm}
\centerline{ X.Q.Yuan$^{a}$, Z.Y.Zhang$^{a}$, Y.W.Yu$^{a}$, P.N.Shen$^{b,a}$}
\vspace{1cm}
{\small
{
\flushleft{\bf  $~~~$a. Institute of High Energy Physics, Chinese Academy
of Sciences,}\\
\flushleft{\bf  $~~~~~~$P.O.Box 918(4),Beijing 100039, China}\\

\vspace{8pt}

\flushleft{\bf  $~~~$b. China Center of Advanced Science and Technology
 (World}\\
\flushleft{\bf  $~~~~~~$Laboratory,P.O.Box 8730, Beijing 100080, China} \\

}}
\vspace{2cm}
\date{}
%\begin{center}
%\begin{minipage}{120mm}
\begin{abstract}
We discuss the structure of Deltaron dibaryon in the chiral $SU(3)$
quark model.
The energy of Deltaron is obtained by considering the coupling of the
$\Delta\Delta$ and $CC$ (hidden color) channels. The effects of
various parameters on the Deltaron mass are also
studied. It is shown that the mass of Deltaron is lower than
the mass of $\Delta\Delta$ but higher than the  mass of $\Delta N \pi$.
\end{abstract}

%\vskip 1in
%\baselineskip 0.1in
%\def\baselinestretch{5.0}
\newpage

Searching dibaryons both theoretically and experimentally has attracted
many physicists' attentions since the first theoretical prediction of
$H$ dibaryon was published by Jaffe  in  1977\cite{s1}. The main reason
is that  because the dibaryon is  a six quark system in a small region
where the one-gluon-exchange and quark exchange play significant roles, it is
a very good place to investigate the quark behavior in the short distance
and to show the new physics on Quantum Chromodynamics (QCD). Up to now,
there were lots of calculations on $H$ dibaryon in various models\cite{s2}.
A current calculation of the  $H$ dibaryon structure by
using the chiral $SU(3)$ quark model \cite{s3} shows that the chiral $SU(3)$
cloud only offers a little attraction to the $H$ dibaryon, and the energy
level of the H state is around the $\Lambda$$\Lambda$
threshold. This result is consistent with  most reports obtained recently
in experiments\cite{s4}. On the other hand,
the chiral $SU(3)$ quark model is quite successful not only in
explaining the structures of baryon ground states but also in
reproducing nucleon-nucleon $(N-N)$ scattering phase shifts and
hyperon-nucleon $(Y-N)$ cross sections\cite{s5}. Thus, in our opinion,
analyzing other possible six-quark states by using such model is
quite meaningful. According to the character of the color-magnetic force
in the one-gluon-exchange interaction,  one found that another interesting
six-quark system, named Deltaron ($S=3,J^{\pi}=3^{+},T=0$), should be studied,
because its color-magnetic character does not have repulsive nature in
comparison with that in the $\Delta\Delta$ state.
In 1987, K.Yazaki \cite{s6} analyzed non-stranged two baryon systems in the
framework of the cluster model.  In his calculation, the one-gluon-exchange
interaction and the confinement potential between two quarks were
considered.  The result showed that in the $NN$, $N \Delta$ and
$\Delta\Delta$ cases, only the $\Delta\Delta$ ($S=3,T=0$) state (Deltaron)
demonstrates attractive feature. Since, $\Delta$ is a resonant
state of nucleon with quite wide width and is easy to decay into $N\pi$,
thus, even the Deltaron is a bound state of $\Delta\Delta$,
it is still not easy
to be measured in the experiment due to its large width, except Deltaron
mass is lower than $NN\pi\pi$ threshold.
F.Wang et.al\cite{s7} studied the structure of the Deltaron state by using
the quark delocalization model.  They found that the Deltaron is a deeply
bound state with a binding energy of 320-390MeV, namely its energy level
is even lower than the threshold of $NN\pi\pi$. However, in their
calculation, the forms of the confinement potentials between all
interacting quark pairs were not a unique form. Those forms depended on
the situation whether these interacting
quarks are in the same "baryon orbit" or in different
"baryon orbits". As we know, in the Generating Coordinate Method (GCM)
framework, the left centered orbit wave function and the right centered
one are not orthogonal to each other, thus, in their calculation,
assuming two different kind confinement potentials in one system might
cause confusion.
  
\vspace{0.5cm}

In this paper, we calculate the energy of Deltaron by solving a coupled
channel equation, $\Delta\Delta$ and $CC$ (hidden color channel), in the
framework of the Resonating Group Method (RGM) with the consideration
of chiral field contributions. As is well known, in the
traditional quark potential model, $V_{qq}$ consists of two parts:
the one-gluon-exchange potential $(OGE)$ governing the
short range interaction and the confinement force dominating mainly the long
range interaction. This simple model achieved
great success in explaining the properties of heavy quarkonia, but met
some problems in light quark systems, especially in studying the $N-N$
force.  One of the problems is the source of the constituent quark mass,
another is lack of medium range attraction in the $N-N$ interaction.
These indicate that part of medium-range non-perturbative QCD
effects is missing in this simple model. Although we do not know how
to derive $V_{qq}$ from the QCD theory vigorously, these problems can
alternatively be treated by introducing constraints of chiral
symmetry which is very important in the  strong interaction\cite{s5}.
In the chiral $SU(3)$ quark model,
starting from the linear expression of the chiral- quark coupling
Lagrangian, we can write the chiral-quark interaction Hamiltonian
as
\begin{equation}
H_{I}^{ch}=g_{ch}\bar{\Psi}(\sum\limits_{a=0}^{8}\sigma_{a}\lambda_{a}+i\sum\limits_{a=0}^{8}\pi_{a}\lambda_{a}\gamma_{5})\Psi,
\end{equation}
where $\sigma_{a}$ denotes four scalar meson fields: $\sigma$,
$\sigma'$, $\kappa$, and $\epsilon$, respectively, and $\pi_{a}$ denotes four pseudoscalar
meson fields: $\eta_{1}$, $\pi$, $K$, and $\eta_{8}$, respectively,
and consequently, the interaction between quarks as
\begin{equation}
V_{ij}^{ch}=V_{ij}^{PS}+V_{ij}^{S},
\end{equation}
with
\begin{equation}
V_{ij}^{PS}=C(g_{ch},m_{\pi_{a}},\Lambda)\frac{m_{\pi_{a}}^{2}}{12m_{i}m_{j}}
[f_{1}(m_{\pi_{a}},\Lambda,r_{ij})(\stackrel{\rightarrow}{\sigma}_{i}\cdot\stackrel{\rightarrow}{\sigma}_{j})+f_{2}(m_{\pi_{a}},\Lambda,r_{ij})S_{ij}](\lambda_{i}^{a}\lambda_{j}^{a})_{f},
\end{equation}
\begin{equation}
V_{ij}^{S}=-C(g_{ch},m_{\sigma_{a}},\Lambda)
%[
f_{3}(m_{\sigma_{a}},\Lambda,r_{ij})
%+\frac{m_{\sigma_{a}}^{2}}{4m
%_{i}m_{j}}f_{4}(m_{\sigma_{a}},\Lambda.r_{ij})
%\stackrel{\rightarrow}{L}\cdot(\stackrel{\rightarrow}{\sigma}_{i}+
%\stackrel{\rightarrow}{\sigma}_{j})]
(\lambda_{i}^{a}\lambda_{j}^{a})_{f},
\end{equation}
here $V_{ij}^{PS}$ and $V_{ij}^{S}$ are the pseudoscalar- and scalar-field
induced interactions respectively.
The expressions of $f_{i}$,$Y$,$G$,$H$,and $C$ are  shown in Ref.[5]
In expressions (3) and (4), there is only one coupling constant $g_{ch}$,
which can be fixed by the following relation:
\begin{equation}
\frac{g_{ch}^{2}}{4\pi}=\frac{g_{NN\pi}^{2}}{4\pi}\frac{9}{25}\frac{m_{q}^{2}}{M_{N}^{2}}.
\end{equation}

Then, the Hamiltonian of the chiral $SU(3)$ quark model reads,
\begin{equation}
H=\sum\limits_{i}T_{i}-T_{G}+\sum_{i<j}V_{ij},
\end{equation}
where $V_{ij}$ includes one-gluon-exchange interaction, confinement potential
and chiral-quark field coupling induced interactions,
\begin{equation}
V_{ij}=V_{ij}^{OGE}+V_{ij}^{conf}+V_{ij}^{ch}.
\end{equation}

\noindent
In Equation (7), $V_{ij}^{OGE}$ is taken in the usual form and
$V_{ij}^{conf}$ is chosen to be  a quadratic form used in Ref.[5]
\begin{equation}
V_{ij}^{conf}=-(\lambda_{i}^{a}\lambda_{j}^{a})_{c}(a_{ij}r_{ij}^{2}
+a_{ij0}).
\end{equation}

\noindent
The coupling constant of $OGE$ and the strength of confinement
potential are determined by the stability condition of nucleon and
the mass difference between $\Delta$ and $N$\cite{s3,s5}.

From the analysis in Ref.[8], we know that the model space of a
simple $(0s)^{6}$ six-quark-cluster configuration is not large enough
to describe the dibaryon structure sufficiently. It is  more effective to
choose the two-cluster configuration as the dibaryon's model space.
According to Ref.[9], the basis of two clusters (i.e physical basis)  and
the six-quark cluster (i.e. symmetry basis) for the case
of $S=3, T=0$ have certain relation. We show it in Table 1.
%\vspace{0.4cm}
\begin{table}
\begin{center}
\caption{Coefficients between the physical basis states}
\centerline{and the symmetry basis states}
\vspace{0.4cm}
\tabcolsep 0.5in
\begin{tabular}{|c|c|c|}\hline
  & $[6][33]_{30}$ & $[42][33]_{30}$ \\\hline
  $(\Delta\Delta)_{30}$ &$\sqrt{\frac{1}{5}}$  &$\sqrt{\frac{4}{5}}$ \\\hline
  $(CC)_{30}$  &$\sqrt{\frac{4}{5}}$    &$-\sqrt{\frac{1}{5}}$  \\\hline
\end{tabular}
\end{center}
\end{table}
In the table, $CC$ channel is the hidden color state which has 
the  form of,
\begin{equation}
\mid CC\rangle_{S=3,T=0}=-\frac{1}{2}\mid\Delta\Delta\rangle_{S=3,T=0}
+\frac{\sqrt{5}}{2}A_{STC}\mid\Delta\Delta\rangle_{S=3,T=0},
\end{equation}
where  $A_{STC}$ is the antisymmetrization operator in the
spin-isospin-color space.

In this work, we choose a model space where both $\Delta\Delta$ and $CC$
channels are included. The mixture of the $L=0$ and $L=2$ states
which shows the effects of the tensor forces from $OGE$
and chiral field coupling are also included, namely
two channels with four states, $\Delta\Delta(L=0)$, $\Delta\Delta(L=2)$,
$CC(L=0)$ and $CC(L=2)$ are considered.
The corresponding matrix elements of spin-isospin-color operators
are given in Appendix.

\vspace{0.7cm}

In the coupled channel bound state calculation, one must carefully
eliminate the forbidden states which may spoil the numerical calculation.
In the Deltaron case, there exists a state with the zero
eigenvalue of the normalization operator, $<N>=0$,  because of the Pauli
blocking effect. It reads

\begin{equation}
\mid\Psi\rangle_{forbidden}=\mid\Delta\Delta\rangle-\frac{1}{2}\mid CC\rangle.
\end{equation}

\noindent
To obtain the reliable result, we perform the  off-shell transformation
to  eliminate  those non-physical degrees.

\vspace{0.5cm}

In the calculation, the same set of model parameters which
can reproduce N-N scattering phase shifts and Y-N cross sections\cite{s5}
are employed. We present the calculated the Deltaron energy and
the root-mean-square radius (RMS) in the chiral $SU(3)$ quark model
in Table 2. For comparison, we also show the results with $OGE$ only and
$OGE$ plus $\pi$, $\sigma$ fields (namely chiral SU(2) coupling) with four
cases,
$\Delta\Delta(L=0)$, $\Delta\Delta(L=0+L=2)$,
$\Delta\Delta+CC(L=0)$ and $\Delta\Delta+CC(L=0+L=2)$.
\begin{table}
\begin{center}
\caption{Energy and RMS of Deltaron dibaryon$^*$}
\vspace{0.4cm}
\tabcolsep 0in
%\begin{small}
\begin{tabular}{|c|c|c|c|c|c|} \hline
\multicolumn{2}{|c|}{}&$\Delta \Delta(L=0)$~  &~$\Delta \Delta \left( \begin{array}{c} L=0 \\ +2
  \end{array} \right)$~
  &$\begin{array}{c} \Delta \Delta  \\ CC \\ \end{array} (L=0)$~
    &$\begin{array}{c}  \Delta \Delta  \\ CC \\ \end{array}
    \left( \begin{array}{c} L=0 \\ +2 \end{array} \right) $~ \\ \hline
& {\em $B~(MeV)$}&29.8  &29.9  &41.0  &42.0  \\ \cline{2-6}
OGE&{\em $\bar{R}$(fm)}&0.92   &0.92    &0.87    &0.87    \\ \hline
 &{\em $B~(MeV)$} &50.2  &62.6  &68.6  &79.7  \\ \cline{2-6}
OGE+$\pi, \sigma$&{\em $\bar{R}$(fm)}&0.87    &0.86    &0.84    &0.83    \\  \hline
     &{\em $B~(MeV)$}      &18.4  &22.5  &31.7  &37.3  \\ \cline{2-6}
OGE+SU(3)~ &{\em $\bar{R}$(fm)}&1.01    &1.00   &0.92    &0.92    \\ \hline
\end{tabular}
\end{center}
\flushleft{\bf \footnotesize {$~~~~~~^* B=-(E_{Deltaron}-2M_{\Delta})$}}\\
\vspace{-0.3cm}
\flushleft{\bf \footnotesize {$~~~~~~~ \bar{R}=\sqrt{<r^2>}$}}
\end{table}
\noindent
From Table 2, we see that the energy of Deltaron is indeed lower than
two ${\Delta}$s', but it is not a deeply bound state. Its binding
energy is always several tens $MeV$ in all three different cases:
the $OGE$, $OGE+SU(2)$ and $OGE+SU(3)$ cases. Since the Deltaron
energy is still higher than the mass of N$\Delta\pi$, it is not a narrow
width dibaryon.

\vspace{0.7cm}

The channel coupling effect is more significant than the L state
mixing effect due to the tensor interaction.  The energy of Deltaron in
the case of $OGE+SU(2)$ is the lowest one.  This means that the $\pi$
and $\sigma$ chiral fields can offer attractions and make the Deltaron
more bound against  $\Delta\Delta$.  However, the total effect of
$SU(3)$ chiral fields would not be able to make the Deltaron energy
lower further.

\vspace{0.7cm}

On the other hand, we investigate the influence of parameter values
on the Deltaron energy. We first examine the effect from the mass of
$\sigma$. In general, the $\sigma$ mass can be
estimated according to the following relation \cite{s10}
\begin{equation}
m_{\sigma}^{2}=(2m_{q})^{2}+m_{\pi}^{2}.
\end{equation}
Therefore, taking $m_{\sigma}$ to be $600-700 MeV$ is reasonable. In Table 2, we
choose $m_{\sigma}=625 MeV$, which is the same as that used in Ref.[5].
For comparison, the results with $m_{\sigma}=550 MeV$ which is a
case in limit are also tabulated in Table 3.
From this table, one sees that the results do not change much even
$m_{\sigma}$ is reduced to a smaller value. Then, we study the
influence of the baryon size
parameter $b_N$ which greatly affects the confinement strength.
In the N-N scattering calculation, $b_N$ is chosen to be $0.505fm$,
the corresponding confinement strength is $a_{c}=54.34 MeV/fm^{2}$ \cite{s5}.
Here, we take another set of parameters where $b_{N}=0.60fm$ and
corresponding $a_{c}=8.19 MeV/fm^{2}$ to show the effect from the
confinement strength. The result is also given in Table 3. It is noted that
the influence from different $b_{N}$ is small.

\begin{table}
\begin{center}
\caption{Deltaron energy B(MeV) with  different parameters$^*$}
\vspace{0.4cm}
\tabcolsep 0.2in
%\begin{small}
\begin{tabular}{|c|c|c|c|c|} \hline
%  &\multicolumn{4}{|c|}{$B ~(MeV)$}  \\ \cline{2-5}
  &\multicolumn{4}{|c|}{$(\Delta\Delta+CC)~~(L=0+L=2)$}  \\ \hline
%OGE  & 41.99  &55.64  &41.99  &55.64  \\ \hline
OGE+$\pi, \sigma$  &79.7  & 97.1 &97.9  &113.4  \\ \hline
OGE+SU(3)~  &37.3  &64.2  &52.4  &79.2  \\ \hline
$b_{N}(fm)$& 0.505 &0.60& 0.505  &0.60 \\ \hline
$m_{\sigma}(MeV)$& 625 &  625  & 550  & 550 \\ \hline
\end{tabular}
%\end{small}
\end{center}
\flushleft{\bf \footnotesize {$~~~~~~^*B=-(E_{Deltaron}-2M_{\Delta})$ }}
\end{table}

As we know, the form of the confinement potential and the
confinement strength only slightly affect the calculated scattering and
bound state results in the two-color-singlet-cluster system.
Now, to study  the Deltaron structure, we have to include the hidden-color
channel $CC$ to enlarge the model space. However, once the $CC$
channel is considered, there would exists the color Van der Waals force
problem. To solve this puzzle, the authors in Ref[11] used an
error-function-like confinement potential to account for the color
screening effect, namely the non-perturbative QCD effect.
Therefore, it is necessary to examine whether the hidden-color state in
the bound state calculation is sensitive to the form of the confinement
potential. For this  purpose, we also adopt the  error-function-like
confinement potential

\begin{equation}
V_{ij}^{erf-conf}=-(\lambda_{i}\cdot\lambda_{j})_{c}
(a_{ij0}+a_{ij}erf(\frac{r}{csl})),
\end{equation}
where $csl$ is the color screening length taken to be $2.0fm$
to calculate the Deltaron structure. The results are shown in Table 4.
From the table, one sees that the energy
and RMS of Deltaron are quite similar to those in
the quadratic confinement case, namely these bound state properties vary
not much as the color screening effect is accounted. The stabilities of
these results reflect the reliability of our calculation.

\begin{table}
\begin{center}
\caption{Energy and RMS of Deltaron with
error-function-like confinement$^*$ }
\vspace{0.4cm}
\tabcolsep 0in
%\begin{small}
\begin{tabular}{|c|c|c|c|c|c|} \hline
\multicolumn{2}{|c|}{}&$\Delta \Delta(L=0)$~  &~$\Delta \Delta \left( \begin{array}{c} L=0 \\ +2
  \end{array} \right)$~
  &$\begin{array}{c} \Delta \Delta  \\ CC \\ \end{array} (L=0)$~
    &$\begin{array}{c}  \Delta \Delta  \\ CC \\ \end{array}
    \left( \begin{array}{c} L=0 \\ +2 \end{array} \right) $~ \\ \hline
OGE  &{\em $B~(MeV)$}&20.9  &21.0  &38.1  &39.1  \\ \hline
%OGE&{\em $\bar{R}$(fm)}& 0.96  &0.96    &0.88     &0.88   \\  \hline
OGE+$\pi,\sigma$  &{\em $B~(MeV)$}  &44.4  &56.8 &71.4  &81.8  \\ \hline
%OGE+$\pi, \sigma$& {\em $\bar{R}$(fm)} &0.88& 0.87  & 0.83  &0.82   \\  \hline
OGE+SU(3)~&{\em $B~(MeV)$}  &13.3  &17.5  &32.5 &38.1  \\ \hline
%OGE+SU(3)~ &{\em $\bar{R}$(fm)}& 1.05& 1.04 &0.92 &0.91\\ \hline
\end{tabular}
%\end{small}
\end{center}
\flushleft{\bf \footnotesize {$~~~~~~^* B=-(E_{Deltaron}-2M_{\Delta})$}}\\
\vspace{-0.3cm}
\flushleft{\bf \footnotesize {$~~~~~~~ \bar{R}=\sqrt{<r^2>}$}}
\end{table}

We finally conclude that in  the framework  of the chiral $SU(3)$ quark
model, the binding energy of the Deltaron is stably ranged
around several tens $MeV$ even  we vary the mass of $\sigma$ and the baryon
size parameter in the reasonable regions and change the form of the
confinement to the error-function-like form to include
the color screening effect, except in the case of OGE+$\pi, \sigma$ with
$b_N=0.6fm$ and $m_{\sigma}=550MeV$, where the binding energy of Deltaron
is about 113$MeV$.  This means that the mass of Deltaron is always
lower than that of $\Delta\Delta$, but higher than that of $\Delta$N$\pi$.
As a conclusion, we see that although the model space has been enlarged,
the color screening effect has been considered and the contributions from
various chiral fields have been included, the binding energy of Deltaron
is always remained around several tens $MeV$. Thus, we announce that
there does not exist a deeply bound Deltaron
dibaryon state with narrow width.

\par

\newpage
\section*{APPENDIX}

\begin{small}

\vspace{0.4cm}
\center{
\renewcommand\arraystretch{0.6}

\tabcolsep 0in
%\centerline{Coefficients of spin-flavor-color operators}
\begin{table}[h]
\caption{Coefficients of  spin-flavor-color operators}
\vspace{0.5cm}
\begin{tabular}{|c|c|c|c|c|c|c|c|c|c|} \hline
\multicolumn{2}{|c|}{$~$} & & & &
\multicolumn{2}{|c|}{$~$} & & &   \\
\multicolumn{2}{|c|}{$~$}
 &$~ \Delta \Delta~ $&  $~\Delta \Delta~ $ & $~$CC$~$&
\multicolumn{2}{|c|}{$~$}  &  $~ \Delta \Delta ~$ &  $ ~\Delta \Delta ~$
&$~$CC$~$ \\
\multicolumn{2}{|c|}{$\hat{O}_{ij}$} & & & &
\multicolumn{2}{|c|}{$\hat{O}_{ij}$} & & &   \\
\multicolumn{2}{|c|}{$~$}  & $ \Delta \Delta $& CC  & CC&
\multicolumn{2}{|c|}{$~$}  &  $ \Delta \Delta $ & CC&  CC
\\
\multicolumn{2}{|c|}{$~$} & & & &
\multicolumn{2}{|c|}{$~$} & & &   \\ \hline
& 1  &27   &0  &27 & & $\hat{O}_{ij}$ &9 &0 &9 \\
& $P_{36}$  &-3   &-12 &-21 &$\stackrel{\rightarrow}{\sigma_{i}}\cdot\stackrel{\rightarrow}{\sigma_{j}}$
& $\hat{O}_{ij}P_{36}$ &-1 &-4 &-7
\\\hline
& $\hat{O}_{12}$ & -72  & 0 & -18  & & $\hat{O}_{12}$ & 9  &0 & -9 \\
& $\hat{O}_{36}$ & 0  & 0 & -36
&
&$\hat{O}_{36}$ & -15  &0 & -3  \\
$\lambda_{i}^c\cdot\lambda_{j}^c$& $\hat{O}_{12}P_{36}$ & 8  & 32 & 2 &
%$~\stackrel{\rightarrow}{\tau_{i}}\cdot\stackrel{\rightarrow}{\tau_{j}}$
$\sum_{k=1}^{3}\lambda^{F}_{i}(k)\lambda^{F}_{j}(k)$
& $\hat{O}_{12}P_{36}$
& -1 &-4 & 11\\
and& $\hat{O}_{13}P_{36}$ & 8  & 32 & 20
&and  &$\hat{O}_{13}P_{36}$  &-1  &-4  &5
\\
$~(\stackrel{\rightarrow}{\sigma_{i}}\cdot\stackrel{\rightarrow}{\sigma_{j}})
(\lambda_{i}^c\cdot\lambda_{j}^c)$~ & $\hat{O}_{16}P_{36}$ & 8 & -4 & 20 &
~$(\stackrel{\rightarrow}{\sigma_{i}}\cdot\stackrel{\rightarrow}{\sigma_{j}})
%(\vec{\tau}_{i} \cdot \vec{\tau}_{j})
(\sum_{k=1}^{3}\lambda^{F}_{i}(k)\lambda^{F}_{j}(k))
$ & $\hat{O}_{16}P_{36}$ & -1  &8 & 5\\
& $\hat{O}_{14}P_{36}$ & -4 & 2 & 35 &   & $\hat{O}_{14}P_{36}$ & 3  &6 & 0  \\
& $\hat{O}_{36}P_{36}$ & -16 & 8 & 32 &   & $\hat{O}_{36}P_{36}$ & 7 & 4 & 1
\\\hline
\multicolumn{2}{|c|}{factor} &~$ \frac{1}{27}$ &~$\frac{1}{27}$ &~$ \frac{1}{27}$ &
\multicolumn{2}{|c|}{factor}  &~$\frac{1}{9}$ &~$\frac{1}{9}$
&~$\frac{1}{9}$~ \\ \hline
\end{tabular}
\end{table}
\newpage
\vspace{0.4cm}

\renewcommand\arraystretch{0.6}
\tabcolsep 0in
\begin{table}
\caption{Coefficients of spin-flavor-color operators(tensor part)}
%\tabcolsep 0in
%\begin{small}
\vspace{0.4cm}
\begin{tabular}{|c|c|c|c|c|c|c|c|c|c|} \hline
\multicolumn{2}{|c|}{$~$} & & & &
\multicolumn{2}{|c|}{$~$} & & &   \\
\multicolumn{2}{|c|}{$~$}
 &$ \Delta \Delta $&  $\Delta \Delta $ &  CC&
\multicolumn{2}{|c|}{$~$}  &  $ \Delta \Delta $ &  $ \Delta \Delta $& CC \\
\multicolumn{2}{|c|}{$\hat{O}_{ij}$} & & & &
\multicolumn{2}{|c|}{$\hat{O}_{ij}$} & & &   \\
\multicolumn{2}{|c|}{$~$}  & $ \Delta \Delta $& CC  & CC&
\multicolumn{2}{|c|}{$~$}  &  $ \Delta \Delta $ & CC&  CC
\\
\multicolumn{2}{|c|}{$~$} & & & &
\multicolumn{2}{|c|}{$~$} & & &   \\  \hline
$ (\stackrel{\rightarrow}{\sigma_{i}}\stackrel{\rightarrow}{\sigma_{j}})_{2}$& $\hat{O}_{ij}$& 54  &0 & 54 & & & & & \\
&$\hat{O}_{ij} P_{36}$   &-6   &-24  &-42 & & & & & \\\hline
& $\hat{O}_{12}$ & -144  & 0 & -36  & & $\hat{O}_{12}$ & 18  &0 & -18 \\
& $\hat{O}_{36}$ & 0  & 0 & -72  &  & $\hat{O}_{36}$ & -30  &0 & -6  \\
& $\hat{O}_{12}P_{36}$ & 16  & 64 & 4 &  & $\hat{O}_{12}P_{36}$ & -2  &-8 & 22\\
$(\stackrel{\rightarrow}{\sigma_{i}}\stackrel{\rightarrow}{\sigma_{j}})_{2}
  (\lambda_{i}^c\cdot\lambda_{j}^c)$ & $\hat{O}_{13}P_{36}$ & 16  & 64 & 40
&$(\stackrel{\rightarrow}{\sigma_{i}}\stackrel{\rightarrow}{\sigma_{j}})_{2}

%  (\vec{\tau}_{i} \cdot \vec{\tau}_{j})
(\sum_{k=1}^{3}\lambda^{F}_{i}(k)\lambda^{F}_{j}(k))

$  &$\hat{O}_{13}P_{36}$  &-2  &-8  &10 \\
& $\hat{O}_{16}P_{36}$ & 16  & -8 & 40 &   & $\hat{O}_{16}P_{36}$ & -2  &16 & 10\\
& $\hat{O}_{14}P_{36}$ & -8 & 4 & 70 &   & $\hat{O}_{14}P_{36}$ & 6  &12 & 0  \\
& $\hat{O}_{36}P_{36}$ & -32 & 16 & 64 &   & $\hat{O}_{36}P_{36}$ & 14 & 8
& 2   \\ \hline
\multicolumn{2}{|c|}{factor} &$ \frac{1}{27} \sqrt{\frac{14}{5}}$ &$\frac{1}{27}
 \sqrt{\frac{14}{5}}$ &$ \frac{1}{27} \sqrt{\frac{14}{5}}$
&\multicolumn{2}{|c|}{factor}  &$\frac{1}{9} \sqrt{\frac{14}{5}}$
&$\frac{1}{9} \sqrt{\frac{14}{5}}$ &$\frac{1}{9} \sqrt{\frac{14}{5}}$ \\ \hline
\end{tabular}
%\end{small}
\end{table}
\center}
\end{small}

\noindent
and

\begin{eqnarray*}
\langle
%\tau_{8}(i) \tau_{8}(j)
\lambda^{F}_{i}(8)\lambda^{F}_{j}(8)
\rangle=\frac{1}{3}\langle 1 \rangle  \\
\langle (\stackrel{\rightarrow}{\sigma_{i}}\cdot \stackrel{\rightarrow}
{\sigma_{j}})(
%\tau_{8}(i) \tau_{8}(j)
\lambda^{F}_{i}(8)\lambda^{F}_{j}(8)
) \rangle
=\frac{1}{3}\langle \stackrel{\rightarrow}{\sigma_{i}}\cdot
\stackrel{\rightarrow}{\sigma_{j}}\rangle   \\
\langle (\stackrel{\rightarrow}{\sigma_{i}}\stackrel{\rightarrow}
{\sigma_{j}})_{2}(
%\tau_{8}(i) \tau_{8}(j)
\lambda^{F}_{i}(8)\lambda^{F}_{j}(8)
) \rangle=
\frac{1}{3}\langle( \stackrel{\rightarrow}{\sigma_{i}}
\stackrel{\rightarrow}{\sigma_{j}})_{2}\rangle
\end{eqnarray*}

\end{document}